# Estimating the Evolution of Solution Norms in Vector Delay Nonlinear Systems: Stability and Boundedness

Mark A. Pinsky
Department of Mathematics and Statistics, University of Nevada. Reno, Reno NV 89557, USA, *Email*: pinsky@unr.edu

**Abstract.** Assessing the boundedness and stability of vector nonlinear systems with variable delays and coefficients remains a challenging problem with broad applications in science and engineering. Existing methods tend to produce overly conservative criteria that offer limited practical value and often fail to explicitly characterize the temporal evolution of solution norms.

This paper presents a novel framework for evaluating the evolution of solution norms in such systems. This approach constructs scalar counterparts for the original vector equations. We prove that the solutions to these scalar nonlinear equations, which also include delays and variable coefficients, provide upper bounds for the norms of the original solutions, if the history functions for both equations are properly matched. This reduction enables the evaluation of the boundedness and stability of vector systems through the analysis of the dynamics of their scalar counterparts, which can be performed via straightforward simulations or simplified analytical reasoning.

Consequently, we introduce new criteria for boundedness and stability and estimate the radii of the balls containing history functions that yield bounded or stable solutions for the original vector systems. Finally, we validated our inferences through representative simulations that also assessed the accuracy of the proposed approach.

## 1. Introduction

Assessing the boundedness and stability of vector nonlinear systems with variable delays and coefficients is a challenging problem that has significant implications in various fields, such as biology, medicine, economics, physics, and control science. As a result, this problem has been extensively studied in numerous contributions, such as monographs [8,9,18,21,22,24-27,29,31-33,37] and surveys [2,3,23,28] along with references therein, where various stability criteria for certain classes of delay systems were developed. The field of delay dynamical systems continues to attract significant new research, which expands its theoretical foundation [5,19,51] and contributes to various applied areas [20, 38,43,45, 52,54].

Key advancements in this field have been made through extensions of the Lyapunov methodology for time-delay systems, initially developed by Krasovskii [31] and Razumikhin [44]. This framework, along with the techniques resting on Halanay's inequality [25] (see also [18, 21, 35]), the Bohl–Perron theorem [4,5, 22], and certain other procedures have shaped the foundation for contemporary advancements in this area.

For linear delay systems, the application of Lyapunov-like techniques has resulted in computationally efficient methods based on linear matrix inequalities (LME), as reviewed in [21, 30, 46].

The stability of systems with time-varying delay systems was also studied in [5,8,9,11,34,54].

However, the application of Lyapunov-like methodology to finite-dimensional time-varying nonlinear systems (ODEs) is challenging already and often leads to overly restrictive stability criteria with conservative estimates of the stability regions. Extending these methods to infinite-dimensional delay nonlinear systems introduces further difficulties, particularly for systems that couple multiple equations with variable delays and coefficients. Consequently, most existing stability criteria for nonlinear delay systems remain overly conservative and rarely offer effective estimates of the stability or attraction regions.

These limitations are even more pronounced when the Lyapunov-like methodology is used to assess the boundedness of solutions to nonhomogeneous vector nonlinear systems with variable delays and coefficients, a topic that has rarely been addressed. Although input-to-state stability analysis has addressed some aspects of the behavior of such systems, as highlighted in a recent survey [3], many challenges remain.

Meanwhile, the concept of finite-time stability, which examines the quantitative and qualitative behavior over bounded time intervals, has also been explored in the context of delay systems; see, for example, a recent survey [53].

Efficient estimation of the stability or boundedness regions for nonlinear, nonautonomous delay systems remains a significant challenge. In principle, these regions can be characterized using level sets of Lyapunov-Krasovskii functionals or Lyapunov-Razumikhin functions, as discussed in [10]–[13], [15],[16], and [47]. However, applying these techniques to vector nonlinear delay systems often results in overly conservative and cumbersome conditions. Attempts to simplify the analysis using the norm of the history function [36], [49] have often made these inferences even more conservative.

On the other hand, fractal structures at the boundaries of stability regions have been reported in [1], [14], and [48].

Most existing techniques do not provide explicit estimates of the temporal evolution of solution norms, which is essential for characterizing boundedness, stability, and instability in dynamical systems.

Building on our previous work on finite-dimensional systems [39-42], this study introduces a novel approach for evaluating the temporal behavior of solution norms in vector nonlinear systems with variable delays and coefficients. The proposed method constructs scalar auxiliary equations that serve as counterparts of the original vector systems. We prove that under a suitable match of history functions, the solutions of these scalar delay nonlinear equations provide upper bounds on the temporal evolution of the solution norms of the original systems.

This framework is the main contribution of this study. It reduces the analysis of the temporal evolution of solution norms in vector nonlinear systems with delays to the study of the dynamics of their scalar counterparts, which are amenable to simplified analytical techniques and efficient simulations. Specifically, we show that, under natural conditions, the solutions of the scalar auxiliary equations depend monotonically on the supremum norm of the history function at each time. This property allows for the formal validation of efficient simulations of the radii of the stability or attraction balls for the original system.

To further streamline the closed-form evaluation of scalar nonlinear delay equations, we bound their solutions by the solutions of linear equations constructed under a Lipschitz-like continuity condition. These equations were then analyzed using the comparison principle.

As a result, we established new criteria for boundedness and stability, derived upper bounds on the norms of the original solutions, and estimated the radii of the history function balls that generate bounded or stable solutions in the original vector systems.

Finally, we validate the proposed approach through representative simulations, which also assess the accuracy of this technique.

The remainder of this paper is organized as follows. The next section introduces the notations, preliminary concepts, and system definitions. Section 3 derives the scalar auxiliary equation and establishes the new criteria for boundedness and stability. Section 4 presents the closed-form robust stability criterion. Section 5 explores the applications of the linearized scalar auxiliary equations. Section 6 reports the simulation results that validate the theoretical analysis. Section 7 concludes this study and outlines potential directions for future research.

## 2. Notation, Mathematical Preliminaries, and Governing Equations

**2.1. Notation**. Firstly, let us recall that the symbols $\mathbb{R}$, $\mathbb{R}_{\geq 0}$, $\mathbb{R}_{+}$ and $\mathbb{R}^n$ represent the sets of real, non-negative, and positive real numbers and real $n$-dimensional vectors. $\mathbb{N}$ is a set of real positive integers, $\mathbb{R}^{n\times n}$ is a set of $n\times n$ matrices, and $I\in\mathbb{R}^{n\times n}$ is an identity matrix. Next, $C([a,b];\mathbb{R}^n)$, $C([a,b];\mathbb{R}_{+})$, and $C([a,b];\mathbb{R}_{\geq 0})$ are the spaces of the continuous real functions $\zeta:[a,b]\to\mathbb{R}^n$, $\zeta:[a,b]\to\mathbb{R}_{+}$ and $\zeta:[a,b]\to\mathbb{R}_{\geq 0}$, respectively, with the supremum norm $\|\zeta\|:=\sup_{t\in[a,b]}|\zeta(t)|$, where $|\cdot|$ stands for the Euclidean norm of a vector or the induced norm of a matrix, $a<b\in\mathbb{R}$ and $b$ can be infinite. In turn, $\dot{x}(t):=D^{+}x(t)$, where $D^{+}$ is the upper-right-hand derivative in $t$. Also note that $|x|_{\infty}=\sup_{i=1,\dots,n}(|x_i|)$ and $|x|_1=\sum_{i=1}^{n}|x_i|$, $\forall x\in\mathbb{R}^n$.

**2.2. Preliminaries.** This section extends the comparison statements for scalar differential equations to scalar delay equations using the method of steps and mathematical induction, under certain additional conditions. Accordingly, we begin by stating two relevant comparison results for scalar ODEs [50].

**Lemma 1**. Consider a scalar equation,

$$\dot{u}=f(t,u),\ u(t_0)=u_0 \tag{2.1}$$

where the function $f\in C([t_0,\infty)\times\mathbb{R};\mathbb{R})$. Next, we assume that (2.1) admits a unique solution, $u(t,u_0)\in J_1\subset\mathbb{R}$, $\forall t\geq t_0$, and we write a coupling scalar inequality:

$$\dot{v}\leq f(t,v),\ v(t_0)=v_0\leq u_0 \tag{2.2}$$

where $v(t,v_0)\in J_1$, $\forall t\geq t_0$ is a solution of (2.2). Then,

$$v(t,v_0)\leq u(t,u_0),\ \forall t\geq t_0 \tag{2.3}$$

The next lemma is often derived by applying Lemma 1 [50].

**Lemma 2**. Consider two scalar equations:

$$\dot{u}_1 = f_1(t,u_1),\ u_1(t_0) = u_{10},$$
$$\dot{u}_2 = f_2(t,u_2),\ u_2(t_0) = u_{20},$$

where functions $f_i \in C([t_0,\infty)\times\mathbb{R};\mathbb{R}),\ i=1,2$. Assume that the above equations admit the following unique solutions: $u_i(t,u_{i0}) \in J_2 \subset \mathbb{R},\ i=1,2,\ \forall t\ge t_0,\ f_1(t,u)\le f_2(t,u),\ \forall u\in J_2,\ \forall t\ge t_0$, and $u_{10}\le u_{20}$. Then, $u_1(t,u_{10}) \le u_2(t,u_{20}),\ \forall t\ge t_0$.

Subsequently, using the method of steps, we extend these results, under some additional assumptions, to scalar functional differential equations with multiple variable delays subject to the following conditions:

$$h_i \in C([t_0,\infty);\mathbb{R}_+),\ \max_i \sup_{\forall t\ge t_0} h_i(t) = \bar{h} < \infty,\ \min_i \inf_{\forall t\ge t_0} h_i(t) \ge \underline{h} > 0,\ i=1,...,m \tag{2.4}$$

which ensures the application of time-stepping with, for example, the time-step that at least equals $\underline{h}$. In turn, we assume that the endpoints of consecutive time steps are set as $s_0 = t_0,\ s_k = t_0 + k\bar{h},\ 1\le k\in\mathbb{N}$. This leads to the following statement.

**Lemma 3**. Denote the scalar delay differential equation and coupling inequality as follows:

$$\dot{u} = f(t,u(t),u(t-h_1(t)),...,u(t-h_m(t))),\ \forall t\ge t_0,\ u(t)=\varphi(t),\ \forall t\in[t_0-\bar{h},t_0] \tag{2.4}$$

$$\dot{v} \le f(t,v(t),v(t-h_1(t)),...,v(t-h_m(t))),\ \forall t\ge t_0,\ v(t)=\phi(t),\ \forall t\in[t_0-\bar{h},t_0] \tag{2.5}$$

where $f(t,x_1,...,x_{m+1})\in\mathbb{R}$ is a continuous function for all variables $\forall t\ge t_0$ and $\forall x_i \in J_3\subset\mathbb{R},\ i=1,...,m+1$. In addition, we assume that $f$ is a nondecreasing function in all variables starting from the third, i.e., $\forall x_i\ i=3,...,m+1,\ \phi,\varphi\in C([t_0-\bar{h},t_0];\mathbb{R}),\ \varphi(t)\le\phi(t),\ \forall t\in[t_0-\bar{h},t_0]$, and continuous functions $h_i(t)$ are subject to condition (2.4). Additionally, we assume that equation (2.5) admits a unique solution: $u(t,\varphi)\in J_3$, $\forall t\ge t_0,\ \forall\|\varphi\|\le\bar{\varphi}\in\mathbb{R}_+$. Then:

$$v(t,\phi)\le u(t,\varphi),\ \forall t\ge t_0 \tag{2.6}$$

where $v(t,\phi)$ is a solution to (2.6).

**Proof**. Let us set that $u_k(t) := u(t,\varphi)$ and $v_k(t) := v(t,\phi),\ \forall t\in[s_{k-1},s_k],\ k\ge 1$. Then, in the first step, relations (2.5) and (2.6) yield:

$$\begin{aligned}\dot{u}_1 &= f(t,u_1(t),\varphi(t-h_1(t)),...,\varphi(t-h_m(t))) = f_1^1(t,u_1),\ \forall t\in[t_0,t_0+s_1]\\ \dot{v}_1 &\le f(t,v_1(t),\phi(t-h_1(t)),...,\phi(t-h_m(t))) = f_2^1(t,v_1),\ \forall t\in[t_0,t_0+s_1]\end{aligned} \tag{2.7}$$

Since $\varphi(t)\ge\phi(t),\ \forall t\in[t_0,t_0+s_1]$ and $f$ are non-decreasing in the variables that include delay, we infer that $f_1^1(t,u)\ge f_2^1(t,u),\ \forall t\in[t_0,t_0+s_1],\ \forall u\in J_3$. The latter inequality allows us to write the second equation in (2.8) as $\dot{v}_1\le f_1^1(t,v_1)$, $\forall t\in[t_0,t_0+s_1]$. Then, the application of Lemma 1 to (2.8) grants $v_1(t)\le u_1(t)$, $\forall t\in[t_0,t_0+s_1]$. Let us, in turn, assume that $v_k(t)\le u_k(t),\ \forall t\in[s_{k-1},s_k],\ 1<k\le K$, and owing to continuity conditions, $u_{k-1}(s_{k-1}) = u_k(s_{k-1})$ and $v_{k-1}(s_{k-1}) = v_k(s_{k-1}),\ 2\le k\le K$. Then, at $k := K+1$, relations (2.5) and (2.6) can be written as follows:

$$\begin{aligned}\dot{u}_{k+1}(t) &= f(t,u_{k+1}(t),u_k(t-h_1(t)),...,u_k(t-h_m(t))) = f_1^{k+1}(t,u_{k+1}),\ \forall t\in[s_k,s_{k+1}]\\ \dot{v}_{k+1}(t) &\le f(t,v_{k+1}(t),v_k(t-h_1(t)),...,v_k(t-h_m(t))) = f_2^{k+1}(t,v_{k+1}),\ \forall t\in[s_k,s_{k+1}]\end{aligned} \tag{2.8}$$

Clearly, inequality $f_1^{k+1}(t,u)\ge f_2^{k+1}(t,u),\ \forall t\in[s_k,s_{k+1}],\ \forall u\in J_3$ holds because of the assumption of function $f$ and the induction hypothesis. Then, applying the latter inequality to (2.9) implies that:

$$\dot{v}_{k+1}(t)\le f(t,v_{k+1},u_k(t-h_1(t)),...,u_k(t-h_m(t))) = f_1^{k+1}(t,v_{k+1}),\ \forall t\in[s_k,s_{k+1}] \tag{2.9}$$

Thus, the correspondence between the first equation in (2.9) and equation (2.10) with equations (2.1) and (2.2), respectively, justifies the application of Lemma 1 □

In turn, the extension of Lemma 2 to scalar delay equations is enabled by the following statement.

**Lemma 4**. Let us write two scalar delay equations in the following form:

$$\begin{aligned} \dot{u}_1 &= f_1\left(t, u_1(t), u_1\left(t-h_1(t)\right), ..., u_1\left(t-h_m(t)\right)\right),\ \forall t \ge t_0,\ u_1(t) = \varphi(t),\ \forall t \in \left[t_0 - \bar{h}, t_0\right] \\ \dot{u}_2 &= f_2\left(t, u_2(t), u_2\left(t-h_1(t)\right), ..., u_2\left(t-h_m(t)\right)\right),\ \forall t \ge t_0,\ u_2(t) = \phi(t),\ \forall t \in \left[t_0 - \bar{h}, t_0\right] \end{aligned} \quad (2.10)$$

where the functions $f_i\left(t, x_1, ..., x_{m+1}\right) \in \mathbb{R}$ are continuous in all variables $\forall t \ge t_0$, $\forall x_j \in J_4 \subset \mathbb{R}$, $j = 1, ..., m+1$, $i = 1, 2$, $\phi, \varphi \in C\left(\left[t_0 - \bar{h}, t_0\right]; \mathbb{R}\right)$, and $\varphi(t) \le \phi(t)$, $\forall t \in \left[t_0 - \bar{h}, t_0\right]$. Next, we assume that the above equations admit unique solutions $u_i(t, \varphi) \in J_4$, $\forall t \ge t_0$, $\forall \|\varphi\|$, $\forall \|\phi\| \le \bar{\varphi} \in \mathbb{R}_+$, $i = 1, 2$, and that

$$f_1\left(t, x_1, x_2 ..., x_{m+1}\right) \le f_2\left(t, x_1, \bar{x}_2, ..., \bar{x}_{m+1}\right),\ \forall t \ge t_0,\ \forall x_j \le \bar{x}_j \in J_4 \subset \mathbb{R},\ \ j = 2, ..., m+1 \quad (2.11)$$

Then, $u_1(t, \varphi) \le u_2(t, \phi)$, $\forall t \ge t_0$.

**Proof**. This statement can be proven as the previous lemma. In fact, let us assume that $u_1^k(t) := u_1(t, \varphi)$, $u_2^k(t) := u_2(t, \phi)$, $\forall t \in \left[s_{k-1}, s_k\right]$, $k \ge 1$. Then, we initially write the following equations:

$$\begin{aligned} \dot{u}_1^1 &= f_1\left(t, u_1^1(t), \varphi\left(t-h_1(t)\right), ..., \varphi\left(t-h_m(t)\right)\right) = f_1^1\left(t, u_1^1\right),\ \forall t \in \left[t_0, t_0 + s_1\right] \\ \dot{u}_2^1 &= f_2\left(t, u_2^1(t), \phi\left(t-h_1(t)\right), ..., \phi\left(t-h_m(t)\right)\right) = f_2^1\left(t, u_2^1\right),\ \forall t \in \left[t_0, t_0 + s_1\right] \end{aligned}$$

Consequently, equation (2.12) and condition $\varphi(t) \le \phi(t)$, $\forall t \in \left[t_0 - \bar{h}, t_0\right]$ imply that $f_1^1(t, u) \le f_2^1(t, u)$, $\forall u \in J_4$, $\forall t \in \left[t_0, t_0 + s_1\right]$. Then Lemma 2 yields that $u_1^1(t) \le u_2^1(t)$, $\forall t \in \left[t_0, t_0 + s_1\right]$. Consequently, we assume that $u_1^k(t) \le u_2^k(t)$, $\forall t \in \left[s_{k-1}, s_k\right]$, $2 \le k \le K$, and that due to continuity $u_i^{k-1}\left(s_{k-1}\right) = u_i^k\left(s_{k-1}\right)$, $2 \le k \le K$, $i = 1, 2$. Next, let us show that the latter inequality can be extended to $k := K+1$. To this end, we write:

$$\begin{aligned} \dot{u}_1^{k+1} &= f_1\left(t, u_1^{k+1}(t), u_1^k\left(t-h_1(t)\right), ..., u_1^k\left(t-h_m(t)\right)\right) = f_1^{k+1}\left(t, u_1^{k+1}\right),\ \forall t \in \left[s_k, s_{k+1}\right], \\ \dot{u}_2^{k+1} &= f_2\left(t, u_2^{k+1}(t), u_2^k\left(t-h_1(t)\right), ..., u_2^k\left(t-h_m(t)\right)\right) = f_2^{k+1}\left(t, u_2^{k+1}\right),\ \forall t \in \left[s_k, s_{k+1}\right] \end{aligned} \quad (2.12)$$

and conclude, owing to both (2.12) and the induction's conjecture, that $f_1^{k+1}(t, u) \le f_2^{k+1}(t, u)$, $\forall u \in J_4$, $\forall t \in \left[s_k, s_{k+1}\right]$, which together with Lemma 2, ensures this statement □

### 2.3. Underlined Equation and Definitions.

In the following, we study the behavior of the solutions to the following vector nonlinear equation with multiple variable delays and coefficients:

$$\dot{x} = A(t)x + f\left(t, x(t), x\left(t-h_1(t)\right), ..., x\left(t-h_m(t)\right)\right) + F(t),\ \forall t \ge t_0,\ x(t, \varphi) = \varphi(t),\ \forall t \in \left[t_0 - \bar{h}, t_0\right] \quad (2.13)$$

where $x(t, \varphi) \in \Re \subset \mathbb{R}^n$, $0 \in \Re$, $f\left(t, \chi_1, ..., \chi_{m+1}\right) \in \mathbb{R}^n$, $\forall t \ge t_0$, $\forall \chi_i \in \mathbb{R}^n$, $i = 1, ..., m+1$ is a continuous function in all variables, $f(t, 0) = 0$, the initial function $\varphi \in C\left(\left[t_0 - \bar{h}, t_0\right]; \mathbb{R}^n\right)$, $\|\varphi\| := \sup_{t \in \left[t_0 - \bar{h}, t_0\right]} \left|\varphi(t)\right|$, scalar functions $h_i(t)$ are defined by (2.4), matrix $A \in C\left(\left[t_0, \infty\right); \mathbb{R}^{n \times n}\right)$ is continuously differentiable, $F(t) = F_0 e(t)$, $e \in C\left(\left[t_0, \infty\right); \mathbb{R}^n\right)$, $\|e\| = \sup_{t \ge t_0} \left|e(t)\right| = 1$, and $F_0 \in \mathbb{R}_{\ge 0}$.

Furthermore, we assume that the initial problem (2.14) admits a unique solution $\forall t \ge t_0$, and $\forall \|\varphi\| \le \bar{\phi} > 0$.

Under these assumptions, the right side of (2.14) is a continuous vector function in all variables, which implies that a solution to this equation, $x\left(t, t_0, \varphi\right)$, is continuous and continuously differentiable in $t$, $\forall t \ge t_0$ and, thus, is bounded on any finite time interval. Consequently, this study primarily focuses on the behavior of the solutions to equation (2.14) as $t \to \infty$.

Note that in what follows, we mostly use the abridged notation for the solutions to (2.14), that is, $x(t,\varphi):=$ $x(t,t_0,\varphi)$, $\forall t\ge t_0$.

To simplify further referencing, we introduce the homogeneous counterpart to (2.14):

$$\dot{x}=A(t)x+f\left(t,x(t),x\left(t-h_1(t)\right),...,x\left(t-h_m(t)\right)\right),\ \forall t\ge t_0,\ x(t,\varphi)=\varphi(t),\ \forall t\in\left[t_0-\bar{h},t_0\right] \qquad (2.14)$$

and the following linear equation:

$$\dot{x}=A(t)x,\ \forall t\ge t_0,\ x(t_0,\varphi)=\varphi(t_0)\in\mathbb{R}^n \qquad (2.15)$$

Subsequently, we assume that (2.15) possesses a unique solution $\forall t\ge t_0$, $\forall\|\varphi\|\le\bar{\phi}$, and write the solution to equation (2.16) as follows: $x(t)=W(t,t_0)\varphi(t_0)$, where $W(t,t_0)=\ w(t)w^{-1}(t_0)$ is the transition (Cauchy) matrix, and $w(t)$ is a fundamental solution matrix for (2.16). We then assume that $|w(t_0)|=1$, a condition that is always satisfiable.

Next, we present the standard definitions of the boundedness and stability of the solutions to equations (2.14) and (2.15), which will be used in the next section of this paper; see, for example,[30].

**Definition 1**. Assume that $\varphi\in C\left(\left[t_0-\bar{h},t_0\right];\mathbb{R}^n\right)$ and equation (2.15) admits a unique solution, $\forall\|\varphi\|\le\bar{\phi}>0$. Then, the trivial solution to equation (2.15) is called:

1) Stable for the set value of $t_0$ if $\forall\varepsilon\in\mathbb{R}_+$, $\exists\delta_1(t_0,\varepsilon)\in\mathbb{R}_+$ such that $|x(t,t_0,\varphi)|<\varepsilon,\ \forall t\ge t_0$ if $\|\varphi\|<\delta_1(t_0,\varepsilon)$. Otherwise, the trivial solution is called unstable.

2) Uniformly stable if in the above definition $\delta_1(t_0,\varepsilon)=\delta_2(\varepsilon)$.

3) Asymptotically stable if it is stable for given value of $t_0$ and $\exists\delta_3(t_0)\in\mathbb{R}_+$ such that $\lim_{t\to\infty}|x(t,t_0,\varphi)|=0$ if $\|\varphi\|<\delta_3(t_0)$.

4) Uniformly asymptotically stable if it is uniformly stable and in the previous definition $\delta_3(t_0)=\delta_4=const$.

5) Uniformly exponentially stable if $\exists\delta_5\in\mathbb{R}_+$ and $\exists c_i\in\mathbb{R}_+$, $i=1,2$ such that:

$$|x(t,t_0,\varphi)|\le c_1\|\varphi\|\exp\left(-c_2(t-t_0)\right),\ \forall t\ge t_0 \text{ if } \forall\|\varphi\|\le\delta_5.$$

**Definition 2**. Assume that equation (2.14) admits a unique solution: $\forall\|\varphi\|\le\bar{\phi}$, $\varphi\in C\left(\left[t_0-\bar{h},t_0\right];\mathbb{R}^n\right)$. The solution to equation (2.14) is called:

6) Locally bounded for the fixed value of $t_0$ if $\exists\delta_6(t_0)\in\mathbb{R}_+$ and $\exists\varepsilon_*(\delta_6)\in\mathbb{R}_+$ such that a condition $\forall\|\varphi\|<\delta_6(t_0)$ implies that $|x(t,t_0,\varphi)|<\varepsilon_*,\ \forall t\ge t_0$.

7) Uniformly bounded if $\delta_6(t_0)=\delta_7=const$.

Furthermore, let $R_i\equiv\sup\delta_i$, $i=3,...,7$ be the superior value of $\delta_i$ for which either the $i$-th condition in Definition 1 or conditions 6 or 7 in Definition 2 hold. This implies that $R_3=R_3(t_0)$, $R_6=R_6(t_0)$. Next, let $B_{R_i}:=\{\varphi\in C\left(\left[t_0-\bar{h},t_0\right],\mathbb{R}^n\right):\|\varphi\|\le R_i\}$, $i=3,...,7$ be the balls with radius $R_i$ centered at the origin. Clearly, these balls encompass history functions that originate bounded and stable solutions to (2.14) and (2.15), respectively.

Next, we recall a definition of robust stability, termed stability under persistent perturbations by Krasovskii [31], pp.161-164]. Consider a perturbed equation for (2.15) in the following form:

$$\begin{aligned}&\dot{x}=A(t)x+f\left(t,x,x\left(t-h_1(t)\right),...,x\left(t-h_m(t)\right)\right)+R\left(t,x,x\left(t-h_1^*(t)\right),...,x\left(t-h_1^*(t)\right)\right),\forall t\ge t_0,\\&x(t)=\varphi(t),\ \forall t\in\left[t_0-\bar{h},t_0\right]\end{aligned} \qquad (2.16)$$

Note that in addition to the conditions set above for the components of (2.17) that are included in (2.15), we also assume the following: (I) function $f(t,\chi_1,...,\chi_{m+1})$ is Lipschitz continuous in $\chi_i\in\mathbb{R}^n$, $i=1,...,m+1$, (II) $f(t,0)=0$, (III) $R\in\mathbb{R}^n$ is a continuous function in all its variables, but $R(t,0)$ might not be zero, and (IV) $h_i^*\in C\left([t_0,\infty);\mathbb{R}_+\right)$ complies with the conditions in (2.4).

**Definition 3**. The trivial solution to equation (2.15) is called robustly stable if $\forall\varepsilon\in\mathbb{R}_+$, $\exists\Delta_i(\varepsilon)\in\mathbb{R}_+$, $i=1,2,3$ such that $\left|x(t,\varphi)\right|_\infty\le\varepsilon,\ \forall t\ge t_0$ if $\left|R(t,\chi_1,...,\chi_{m+1})\right|_\infty<\Delta_1(\varepsilon),\ \forall t\ge t_0,\ \forall\left|\chi_i\right|_\infty<\varepsilon,\ i=1,...,m+1$, $\|\varphi\|<\Delta_2(\varepsilon)$, $\left|h_i(t)-h_i^*(t)\right|<\Delta_3(\varepsilon),\ \forall t\ge t_0,\ i=1,...,m$, and $x(t,\varphi)$ are solutions to equation (2.17).

Clearly, this definition defines the robust stability of (2.15) through the boundedness of the solutions to the perturbed equation (2.17).

**Statement** (Krasovskii [31], p.162). Assume that conditions (I)–(IV) hold and that the trivial solution of (2.15) is uniformly asymptotically stable. Then, it is also robustly stable.

Lastly, we extend the definition of finite-time stability, originally introduced for ODEs in [17], to systems with delays.

**Definition 4**. Equation (2.14) and (2.15) are called:

a) Finite-time stable (FTS) with respect to positive numbers $\alpha,\beta,T,\ \alpha<\beta$ if a condition $\|\varphi\|<\alpha$ implies $\left|x(t,t_0,\varphi)\right|<\beta(\alpha,T),\ \forall t\in[t_0,t_0+T]$.

b) Finite-time contractively stable (FTCS) with respect to positive numbers $\alpha,\beta,\gamma,T,\ \alpha<\gamma$ if it is FTS with respect to $\alpha,\beta,T$ and $\exists t_1\in(t_0,t_0+T)$ such that $\left|x(t,t_0,\varphi)\right|<\gamma(\alpha,T),\ \forall t\in[t_1,t_0+T]$.

### 3. Scalar Delay Auxiliary Equations

This section derives a scalar delay auxiliary differential equation and the corresponding history function, which enables the estimation from above of the temporal evolution of the norm of the solutions to the vector delay nonlinear equation (2.14).

First, we write (2.14) in its integral form as follows:

$$x(t,\varphi)=w(t)w^{-1}(t_0)\varphi(t_0)+w(t)\int_{t_0}^{t}w^{-1}(\tau)Q(\tau)\,d\tau,\ \forall t\ge t_0,$$
$$x(t,\varphi)=\varphi(t),\ \forall t\in\left[t_0-\bar{h},t_0\right]$$

where $Q=f\left(t,x(t),x(t-h_1(t)),...,x(t-h_m(t))\right)+F(t)$

The latter equation leads to the following relation:

$$\begin{aligned}&\left|x(t,\varphi)\right|=\left|w(t)w^{-1}(t_0)\varphi(t_0)+w(t)\int_{t_0}^{t}w^{-1}(\tau)Q(\tau)\,d\tau\right|,\ \forall t\ge t_0,\\&\left|x(t,\varphi)\right|=\left|\varphi(t)\right|,\ \forall t\in\left[t_0-\bar{h},t_0\right]\end{aligned}\tag{3.1}$$

Then, the application of the standard norm's inequalities yields the following equation:

$$\begin{aligned}&\left|X_1(t,\varphi)\right|=\left|w(t)\right|\left|w^{-1}(t_0)\varphi(t_0)\right|+\left|w(t)\right|\int_{t_0}^{t}\left|w^{-1}(\tau)\right|\left|Q(\tau)\right|d\tau,\ \forall t\ge t_0,\\&\left|X_1(t,\varphi)\right|=\left|\varphi(t)\right|,\ \forall t\in\left[t_0-\bar{h},t_0\right]\end{aligned}\tag{3.2}$$

where $X_1(t,\varphi)\in\mathbb{R}^n$ is a continuous function in both variables. Comparing (3.1) and (3.2) yields that $\left|x(t,\varphi)\right|\le\left|X_1(t,\varphi)\right|,\ \forall t\ge t_0$.

To write (3.2) in a more tractable form, we introduce a nonlinear extension of the Lipschitz continuity condition, as follows:

$$\left|f(t,\chi_1,...,\chi_{m+1})\right|\le L\left(t,\left|\chi_1\right|,,...,\left|\chi_{m+1}\right|\right),\ \forall\chi=\left[\chi_1,...,\chi_{m+1}\right]^T\in\Omega\in\mathbb{R}^{n(m+1)},\ \chi_i\in\mathbb{R}^n,\ i=2,...,m+1,\tag{3.3}$$

where $\Omega$ is a compact subset of $\mathbb{R}^{n(m+1)}$ containing zero, $f\in C\left([t_0,\infty)\times\mathbb{R}^{n(m+1)};\mathbb{R}^n\right)$, $f(t,0)=0$, $L\in C\left([t_0,\infty)\times\mathbb{R}_{\ge0}^{m+1};\mathbb{R}_{\ge0}\right)$ is a scalar continuous function, and $L(t,0)=0$.

In the Appendix, we illustrate how to define such scalar functions in a closed form if $f(t,\chi_1,...,\chi_{m+1})$ is a polynomial or power series in $\chi_1,...,\chi_{m+1}$ with, for instance, bounded in $\Omega$ error term. In the former case, $\Omega \equiv \mathbb{R}^{n(m+1)}$ and this condition also holds if the error term of the power series is bounded in $\mathbb{R}^{n(m+1)}$. Additionally, $L$ turns out to be a linear function in $|\chi_i|$ if $f$ is a linear function in the corresponding variables. Note also that function $L$ is defined in Section 6 for a planar nonlinear system.

It is also worth noting that in the remainder of this paper we assume, for simplicity, that $\Omega \equiv \mathbb{R}^{n(m+1)}$.

Next, the utility of (3.3) in (3.2) yields the following equation:

$$\begin{aligned} X_2(t,\tilde{\varphi}) &= |w(t)|\left|w^{-1}(t_0)\varphi(t_0)\right| + |w(t)|\int_{t_0}^{t}\left|w^{-1}(\tau)\right| q(\tau)\,d\tau,\ \forall t \ge t_0 \\ X_2(t,\tilde{\varphi}) &= |\varphi(t)|,\ \forall t \in \left[t_0-\bar{h},t_0\right] \end{aligned} \tag{3.4}$$

where $X_2(t,\tilde{\varphi}) \in \mathbb{R}_{\ge 0}$ is a continuous function of both variables and $q = L\left(t, X_2(t), X_2(t-h_1(t)),..., X_2(t-h_m(t))\right) + |F(t)|$. Hence, equation (3.4) implies that $|x(t,\varphi)| \le |X_1(t,\varphi)| \le X_2(t,\tilde{\varphi}),\ \forall t \ge t_0$.

Next, we match (3.4) with the solution to the initial problem of the following scalar-delay differential equation:

$$\begin{aligned} \dot{X}_3 &= p(t)X_3 + c(t)\left(L\left(t, X_3(t), X_3(t-h_1(t)),..., X_3(t-h_m(t))\right) + |F(t)|\right),\ \forall t \ge t_0 \\ X_3(t,\phi) &= \phi(t),\ \forall t \in \left[t_0-\bar{h},t_0\right] \end{aligned} \tag{3.5}$$

where, $X_3(t,\phi) \in \mathbb{R}_{\ge 0}$, $p:[t_0,\infty) \to \mathbb{R}$, $c:[t_0,\infty) \to [1,\infty]$ and $\phi \in C\left(\left[t_0-\bar{h},t_0\right];\mathbb{R}_{\ge 0}\right)$.

To define the functions $p(t)$, $c(t)$, and $\phi(t)$, we write (3.5) in integral form:

$$\begin{aligned} X_3(t) &= \mathrm{e}^{d(t)}\left(\phi(t_0) + \int_{t_0}^{t}\mathrm{e}^{-d(\tau)}c(\tau)q(\tau)\,d\tau\right),\ \forall t \ge t_0 \\ X_{\backslash 3}(t,\phi) &= \phi(t),\ \forall t \in \left[t_0-\bar{h},t_0\right] \end{aligned} \tag{3.6}$$

where $d(t) = \int_{t_0}^{t} p(s)\,ds$.

To determine $p(t)$ and $c(t)$, we match the right-hand sides of (3.6) and (3.4). Matching the first additions in these expressions, that is, $|w(t)|\left|w^{-1}(t_0)\varphi(t_0)\right|$ and $\mathrm{e}^{d(t)}\phi(t_0)$, returns:

$$|w(t)| = \exp\left(\int_{t_0}^{t} p(s)\,ds\right) \tag{3.7}$$

$$\phi(t_0) = \left|w^{-1}(t_0)\varphi(t_0)\right| \tag{3.8}$$

To match the second additions on the right side of (3.6) and (3.4), we first multiply and divide the latter function by $|w(t)|$ and, also, use that $\mathrm{e}^{-d(t)} = 1/|w(t)|$ due to (3.7), which returns:

$$\int_{t_0}^{t}\mathrm{e}^{-d(\tau)}c(\tau)q(\tau)\,d\tau = \int_{t_0}^{t}\left(|w(\tau)|\left|w^{-1}(\tau)\right|\right)q(\tau)/|w(\tau)|\,d\tau$$

The last relation yields that:

$$c(t) = |w(t)|\left|w^{-1}(t)\right|$$

is the running condition number of $w(t)$. Recall that $0 < \left|w^{-1}(t)\right| < \infty,\ \forall t \ge t_0$, which implies $c(t) < \infty,\ \forall t \ge t_0$. Additionally, the continuous differentiability of $A(t)$ implies the corresponding property for $c(t)$ and continuity of $p(t)$.

Lastly, from (3.7) we get that,

$$p(t) = d\left(\ln|w(t)|\right)/dt \tag{3.9}$$

Next, matching the initial functions in (3.4) and (3.6) yields that $\phi(t)=|\varphi(t)|,\ \forall t\in[t_0-\bar{h},t_0]$, which together with (3.8) implies the following condition: $|w^{-1}(t_0)\varphi(t_0)|=|\varphi(t_0)|$. Thus, to satisfy our previous condition, $|w(t_0)|=1$, we set that $w(t_0)=I$ .

Consequently, we write (3.5) as follows:

$$\begin{aligned} &\dot{y}=p(t)y+c(t)\Big(L\big(t,y(t),y(t-h_1(t)),...,y(t-h_m(t))\big)+|F(t)|\Big) \\ &y(t,\phi)=|\varphi(t)|,\ \forall t\in[t_0-\bar{h},t_0] \end{aligned} \tag{3.10}$$

where $y(t)\in\mathbb{R}_{\geq 0}$ . Thus, (3.10) defines the initial problem for the derived scalar auxiliary equation, whose solutions provide an upper bound for the norm of solutions to the initial vector system (2.14). This leads to the following statement.

**Theorem 1**. Assume that $f(t,\chi_1,...,\chi_{m+1})\in\mathbb{R}^n,\ \chi_i\in\mathbb{R}^n,\ i=2,...,m+1$ is a continuous function in all variables and locally Lipschitz in the second one, $f(t,0)=0$ , matrix $A\in C([t_0,\infty);\mathbb{R}^{n\times n})$ is continuously differentiable, $\varphi\in C([t_0-\bar{h},t_0];\mathbb{R}^n)$, scalar and continuous functions $h_i(t)$ are defined by (2.4), $F(t)=F_0e(t)$, $e\in C([t_0,\infty);\mathbb{R}^n)$, $\|e\|=1$, $F_0\in\mathbb{R}_{\geq 0}$, $L(t,\zeta_1,,...,\zeta_{m+1})\in\mathbb{R}_{\geq 0}$ , $\zeta_i\in\mathbb{R}_{\geq 0}$, $i=2,...,m+1$ is a scalar function that is continuous in all variables, $L(t,0)=0$ , inequality (3.3) holds with $\Omega\equiv\mathbb{R}^{n(m+1)}$ , and $w(t_0)=I$ . Subsequently, we assume that equations (2.14) and (3.10) assume unique solutions $\forall\|\varphi\|\leq\bar{\varphi}\in\mathbb{R}_+$, $\forall t\geq t_0$ . Then,

$$\begin{aligned} &|x(t,\varphi)|\leq y(t,\phi),\ \forall t\geq t_0 \\ &\phi(t)=|\varphi(t)|,\ \forall t\in[t_0-\bar{h},t_0] \end{aligned} \tag{3.11}$$

where $x(t,\varphi)$ and $y(t,\phi)\geq 0$ are the solutions to equations (2.14) and (3.10), respectively.

**Proof**. In fact, it has been shown previously that $|x(t,\varphi)|\leq X_2(t,\tilde{\varphi}),\ \forall t\geq t_0$. Assignments on the functions $p(t),\ c(t)$ and $\phi(t)$ made above yield: $X_2(t,\tilde{\varphi})\equiv X_3(t,\phi)\equiv y(t,\phi),\ \forall t\geq t_0$ if $\tilde{\varphi}(t)=\phi(t)=|\varphi(t)|$, $\forall t\in[t_0-\bar{h},t_0]$ □

Therefore, enforcing a boundedness or stability condition on scalar equation (3.10) guarantees that the same property carries over to vector equations (2.14) and (2.15), respectively.

To simplify further referencing, we present the homogeneous counterpart of (3.10) as follows:

$$\begin{aligned} &\dot{y}=p(t)y+c(t)L\big(t,y(t),y(t-h_1(t)),...,y(t-h_m(t))\big) \\ &y(t,\phi)=|\varphi(t)|,\ \forall t\in[t_0-\bar{h},t_0] \end{aligned} \tag{3.12}$$

and assume that (3.12) admits a unique solution under the conditions of Theorem 1.

Next, let us assume that Definitions 1 and 2 are concerned with the solutions to equations (3.12) and (3.10) and that $r_i$ are the superior values of $\delta_i$ , that is, $r_i\equiv\sup\delta_i,\ i=3,...,7$, for which the $i$ -th condition in Definition 1 holds for (3.12) or conditions 6 or 7 in Definition 2 hold for (3.10). Consequently, $r_3=r_3(t_0),\ r_6=r_6(t_0)$. Then, let $B_{r_i}:=\{\varphi\in C([t_0-\bar{h},t_0],\mathbb{R}^n):\ \|\varphi\|\leq r_i\},\ i=3,...,7$ be the balls with radiuses $r_i$ that are centered at the origin.

This leads to the following statements.

**Theorem 2**. Assume that the conditions of Theorem 1 are satisfied and that the trivial solution to equation (3.12) is either stable, uniformly stable, asymptotically stable, uniformly asymptotically stable, or exponentially stable. Then, the trivial solution to equation (2.15) exhibits the same type of stability.

Furthermore, under the conditions 3-5 of Definition 1, $B_{r_i}\subseteq B_{R_i}$, $i=3,4,5$ .

**Proof**. Both above statements follow directly from the application of inequality (3.11) to the corresponding solutions to equations (2.15) and (3.12) □

**Theorem 3**. Assume that the conditions of Theorem 1 are met and that the solutions to equation (3.10) are bounded for the set values of $t_0$ if $\|\varphi\| \le r_6$ or uniformly bounded if $\|\varphi\| \le r_7$. Then, the solutions to equation (2.14) with the matched history functions are also bounded for the same values of $t_0$ or are uniformly bounded. Furthermore, under prior conditions, $B_{r_i} \subseteq B_{R_i}$, $i = 6,7$.

**Proof**. The proof of this statement follows directly from the application of inequality (3.11) to the solutions of equation (2.14) and its scalar counterpart (3.10) □

The key task of estimating the values of $r_i$, $i = 3,...,7$ is abridged by the following statement.

**Lemma 5**. Assume that $y_i(t,\phi_i)$, $\forall t \ge t_0$, $i = 1,2$ are the solutions to (3.10) or (3.12) and $\phi_1(t) \ge \phi_2(t), \forall t \in [t_0 - \bar{h}, t_0]$. Then, $y_1(t,\phi_1) \ge y_2(t,\phi_2)$, $\forall t \ge t_0$.

**Proof**. The proof of this statement follows from the assumption of uniqueness of solutions to equations (3.10) and (3.12), which rules out the intersection of the solution curves to a scalar delay equation in $t \times y$ - plane □

Thus, a solution to equations (3.10) and (3.12) with a constant history function, that is, $y = y(t,\phi)$, $\forall t \ge t_0$, $\phi(t) \equiv q \in \mathbb{R}$, $\forall t \in [t_0 - \bar{h}, t_0]$, is a monotonic function in $q$, $\forall t \ge t_0$, which further streamlines the simulations of the values of $r_i$.

Next, we apply our approach to ensure the robust stability of the trivial solution of the vector delay equation (2.15) via conditions imposed on its scalar counterpart. Consequently, we can write the scalar counterpart for (2.17) as follows:

$$\begin{aligned} &\dot{z} = p(t)z + c(t)\Big(L\big(t, z(t), z(t-h_1(t)),..., z(t-h_m(t))\big) + L_R\big(t, z(t), z(t-h_1^*(t)),..., z(t-h_m^*(t))\big)\Big), \forall t \ge t_0 \\ &z(t,\phi) = |\varphi(t)|, \ \forall t \in [t_0 - \bar{h}, t_0] \end{aligned} \tag{3.13}$$

where $z(t,\phi) \in \mathbb{R}_{\ge 0}$ and $L_R(t,\xi_1,...,\xi_{m+1}) \in \mathbb{R}_{\ge 0}$, $\forall \xi_i \in \mathbb{R}_{\ge 0}$, $\forall t \ge t_0$ is continuous function in all its variables.

Subsequently, the application of Definition 3 to (3.12) and (3.13) leads to the following statement.

**Theorem 4**. Assume that the conditions of Theorem 1 and conditions (I)–(IV) hold for equation (3.13), and that the trivial solution to equation (3.12) is uniformly asymptotically stable. Then, the trivial solution to equation (3.13) and, consequently, to equation (2.15), is robustly stable.

**Proof**. In fact, under the above conditions, the trivial solution of (3.12) is robustly stable which implies that $z(t,|\varphi|) < \varepsilon$, $\forall t \ge t_0$, where $z(t,|\varphi|)$ is a solution to (3.13), if $|L_R(t,\xi_1,...,\xi_{m+1})|_\infty < \Delta_1(\varepsilon)$, $\forall t \ge t_0$, $\forall \xi_i < \varepsilon$, $i = 1,...,m+1$, $\|\varphi\| < \Delta_2(\varepsilon)$, and $|h_i(t) - h_i^*(t)| < \Delta_3(\varepsilon)$, $\forall t \ge t_0$, see Definition 3. Then, (3.11) yields that $|x(t,\varphi)| \le z(t,|\varphi|) < \varepsilon$, $\forall t \ge t_0$, where $x(t,\varphi)$ is a solution to (2.17), since (3.13) is the scalar counterpart to (2.17) □

Lastly, we show that the FTS conditions for equations (2.14) and (2.15) can also be derived by applying the corresponding conditions to their scalar counterparts. This leads to the following statement.

**Theorem 5**. Assume that:

a) Equation (3.10) or equation (3.12) is FTS with respect to $\alpha,\beta,T \in \mathbb{R}_+$, $\alpha < \beta$. Then, equation (2.14) or equation (2.15) is the FTS with respect to $\alpha,\beta,T$, respectively.

b) Equation (3.10) or equation (3.12) is the FTCS with respect to $\alpha,\beta,\gamma,T \in \mathbb{R}_+$, $\alpha < \gamma$. Then, equation (2.14) or equation (2.15) is the FTCS with respect to $\alpha,\beta,\gamma,T$, respectively.

**Proof**. In fact, both above statements directly follow from (3.11) □

Thus, the proposed approach simplifies the analysis of the boundedness and stability of vector nonlinear systems with variable delays and coefficients by enabling rigorously justified and seamless simulations of their scalar counterparts. These simulations also estimate the radii of the history function balls that yield bounded or asymptotically stable solutions to the original system. In addition, this technique simplifies the application of analytical methods to these key problems.

## 4. A Closed-Form Robust Stability Criterion

Analytical criteria for boundedness or stability, initially developed for scalar delay equations (see, e.g., [6] and additional references therein), can often be extended to vector systems using the methodology described above.

Furthermore, simplified versions of equations (3.10) and (3.12) can be obtained by bounding their right-hand sides by time-invariant or linear functions. Applications of these approaches are discussed in the following sections.

The next statement presents a criterion for the robust stability of the trivial solution to a nonlinear, nonautonomous system of ODEs subject to nonlinear perturbations, which include variable delays and coefficients.

**Corollary 1**. Assume that the conditions of Theorem 1 hold, and also $\sup_{\forall t \ge t_0} p(t) = \hat{p} < 0$, $\sup_{\forall t \ge t_0} c(t) = \hat{c} < \infty$, and $\sup_{\forall t \ge t_0} L(t, y) = \hat{L}(y) < \infty$. Next, we presume that $\hat{p}\, y + \hat{c}\hat{L}(y) < 0,\ \forall y \in (0, y_+),\ y_+ > 0$ and conditions (I)–(IV) hold for (3.13). Then, the trivial solution to equation (2.15) is robustly stable.

**Proof**. Under the given assumptions, the trivial solution to scalar equation (3.12) is uniformly asymptotically stable. Due to Theorem 4, this implies that equation (2.15) is robustly stable □

## 5. Linearized Delay Scalar Auxiliary Equations

This section linearizes scalar equations (3.10) and (3.12) near the origin using a Lipschitz-type continuity condition. Combined with the comparison principle for scalar delay equations (Lemma4), this linearization simplifies the criteria for the boundedness and stability of the scalar nonlinear delay equations, and by extension, of their corresponding vector systems.

Let us assume that,

$$L(t, \zeta_1, ..., \zeta_{m+1}) \le \sum_{i=1}^{m+1} \mu_i(t, \tilde{\zeta})\zeta_i,\ \forall |\zeta| \le \tilde{\zeta} > 0,\ \forall t \ge t_0 \tag{5.1}$$

where $\zeta_i \in \mathbb{R}_{\ge 0}$, $\zeta = [\zeta_1, ..., \zeta_{m+1}]^T$ and $\mu_i \in C([t_0, \infty) \times \mathbb{R}_{\ge 0}; \mathbb{R}_{\ge 0})$. Then, the application of (5.1) to (3.10) yields the following scalar linear equation:

$$\begin{aligned} &\dot{u} = P(t)u + c(t)\left(\sum_{i=2}^{m+1} \mu_i(t, \tilde{\zeta}) u(t - h_i(t)) + F_0 |e(t)|\right) \\ &u(t, \phi) = |\varphi(t)|,\ \forall t \in [t_0 - \bar{h}, t_0] \end{aligned} \tag{5.2}$$

where $P(t) = p(t) + c(t)\mu_1(t, \tilde{\zeta})$. To further shorten citations, we also write a homogeneous counterpart to (5.2) as follows:

$$\begin{aligned} &\dot{u} = P(t)u + c(t)\sum_{i=2}^{m+1} \mu_i(t, \tilde{\zeta}) u(t - h_i(t)) \\ &u(t, \phi) = |\varphi(t)|,\ \forall t \in [t_0 - \bar{h}, t_0] \end{aligned} \tag{5.3}$$

Next, we set in (5.1) that,

$$\zeta_1 = u(t, \phi),\ \zeta_i = u(t - h_i(t), \phi),\ i = 2, ..., m+1 \tag{5.4}$$

where $u(t, \phi)$ is a solution to either equation (5.2) or equation (5.3).

In the latter case, we show that inequality (5.1) holds if (5.3) is stable and $\|\varphi\|$ is sufficiently small. In the former case, we show that (5.1) holds under the same conditions, provided that $F_0$ is also sufficiently small.

These observations lead to the following statements.

**Lemma 6**. Assume that $\mu_i \in C([t_0, \infty) \times \mathbb{R}_{\ge 0}; \mathbb{R}_{\ge 0}),\ i = 1, ..., m+1,\ p \in C([t_0, \infty); \mathbb{R}),\ c \in C([t_0, \infty); \mathbb{R}_+)$, $\varphi \in C([t_0, \infty); \mathbb{R}^n)$, and $h_i(t)$ comply with (2.4), equation (5.3) admits a unique solution, $\forall \tilde{\zeta} < \zeta_* > 0$, $\forall t \ge t_0$, and $\forall \|\varphi\| \le \hat{\varphi} > 0$. In addition, we assume that the trivial solution to (5.3) is stable for the set value of $t_0$ and $\forall \tilde{\zeta} \le \tilde{\zeta}_s \le \zeta_*$, where $\tilde{\zeta}_s$ is the supremum value of $\tilde{\zeta}$ for which (5.3) is stable. Then, inequality (5.1) holds if $\|\varphi\|$ is sufficiently small, where $\zeta_i(t, \phi)$ are assigned by (5.4), and $u(t, \phi)$ is a solution to (5.3).

**Proof**. Applying Definition 1a to equation (5.3) implies that under the given conditions, $u(t,\phi)<\varepsilon$, $\forall t\geq t_0$, where $\varepsilon$ is an arbitrarily small positive value if $\|\varphi\|$ is sufficiently small □

The last lemma leads to the following stability criteria.

**Theorem 6**. Assume that the conditions of Theorem 1 and Lemma 6 hold. Then:

$$|x(t,\varphi)|\leq y(t,\phi)\leq u(t,\phi),\ \forall t\geq t_0 \tag{5.5}$$

where $x(t,\varphi)$, $y(t,\phi)$ and $u(t,\phi)$ are the solutions to equations (2.15), (3.12), and (5.3), respectively. Furthermore, the trivial solutions to both equations (3.12) and (2.15) are stable for the set value of $t_0$.

**Proof**. In fact, inequality (5.1) holds under these conditions, because of Lemma 6. This, in turn, ensures that inequality (2.12) holds for equations (3.12) and (5.3). Then, applying Lemma 4 to scalar equations (3.12) and (5.3) yields $y(t,\phi)\leq u(t,\phi)$, $\forall t\geq t_0$, which implies the stability of the trivial solutions to scalar equation (3.12) and, subsequently, to vector equation (2.15) □

**Corollary 2**. Assume that the conditions of Theorem 6 hold and that the trivial solution to (5.3) is either uniformly stable, asymptotically stable, uniformly asymptotically stable, or exponentially stable. Then, the trivial solutions of equations (3.12) and (2.15) exhibit the same type of stability.

**Proof**. The current conditions are, in fact, more conservative than those stated previously. Thus, inequalities (5.1) and (5.5) hold under these conditions, thereby validating the statement, □

To further simplify equation (5.3), we also assume that $\sup_{\forall t\geq t_0} p(t)=\hat{p}<0$, $\sup_{\forall t\geq t_0} c(t)=\hat{c}<\infty$, $\sup_{\forall t\geq t_0} \mu_i(t,\tilde{\zeta})=\hat{\mu}_i(\tilde{\zeta})$, $<\infty$ and review (5.3) as follows:

$$\dot{U}=PU+\hat{c}\left(\sum_{i=2}^{m+1}\hat{\mu}_i(\tilde{\zeta})U(t-h_i(t))\right),\ \ U(t,\phi)=|\varphi(t)|,\ \forall t\in\left[t_0-\bar{h},t_0\right] \tag{5.6}$$

where $P=\hat{p}+\hat{c}\hat{\mu}_1$ and $U\in C([t_0,\infty);\mathbb{R}_{\geq 0})$. This leads to the following statement.

**Corollary 3**. Assume that (5.6) admits a unique solution $\forall\tilde{\zeta}<\zeta_\times\leq\zeta_*$, $\forall t\geq t_0$, and $\forall\|\varphi\|\leq\varphi_\times\leq\hat{\varphi}$. In addition, we assume that (5.6) is stable for the set value of $t_0$ and $\forall\tilde{\zeta}\leq\tilde{\zeta}_U\leq\tilde{\zeta}_s$, and that $\|\varphi\|$ is a sufficiently small value. Then,

$$|x(t,|\varphi|)|\leq y(t,|\varphi|)\leq u(t,|\varphi|)\leq U(t,|\varphi|),\ \forall t\geq t_0$$

and the trivial solutions to equations (3.12) and (2.15) are stable for the set value of $t_0$.

**Proof**. The proof follows the same reasoning as in Lemma 6 and Theorem 6. Specifically, the given conditions allow Lemma 4 to be used for equations (5.3) and (5.6). Consequently, the inequality $u(t,|\varphi|)\leq U(t,|\varphi|)$, $\forall t\geq t_0$ follows from applying Definition 1a and Lemma 4 to these equations □

Thus, the stability of the linear scalar delay equation (5.3) and its autonomous counterpart (5.6) implies the stability of the trivial solution of the vector nonlinear system (2.15) with variable delays and coefficients. The stability criteria for (5.3) and (5.6) can be found in [7] and in the references therein. Moreover, note that the stability criteria for (5.6) are readily known if $h_i\equiv const$, see, for instance, [21, 24, 30].

In turn, we extend this approach to establish sufficient boundedness criteria for equation (2.14) by analyzing the solutions of the linear scalar equation (5.2), which can be represented as follows (see, e.g., [30], p.139):

$$u(t,\phi)=u_h(t,\phi)+F_0u_{nh}(t) \tag{5.7}$$

where $u_h(t,\phi)$ is a solution to (5.3) and $u_{nh}(t)$ is a solution to (5.2) with $\varphi(t)\equiv 0$ and $F_0=1$. In turn, $u_{nh}(t)=\int_{t_0}^{t}C(t,s)|e(s)|ds,\ \forall t\geq t_0$, where $C(t,s)$ is a Cauchy function for (5.2), which is defined as a solution to (5.3) with the following history function: $C(t,s)=\left\{\begin{matrix}0, & t_0-\bar{h}\leq t<s\\ 1, & t=s\end{matrix}\right\}$. This leads to the following statement.

**Lemma 7**. Assume that the conditions of Lemma 6 are satisfied, and equation (5.2) has a unique solution under these conditions. In addition, we assume that $|u_{nh}(t)| \le \hat{u}_{nh} < \infty, \ \forall t \ge t_0$, and both $F_0$ and $\|\varphi\|$ are sufficiently small. Then, inequality (5.1) holds, where $\zeta_i(t,\phi)$ are assigned by (5.4) and $u(t,\phi)$ is a solution to (5.2).

**Proof**. In fact, under the above conditions, both terms on the right-side of (5.7) can be made arbitrarily small $\forall t \ge t_0$ by appropriately choosing $F_0$ and $\|\varphi\|$, which implies that (5.1) holds □

Next, we introduce a criterion for the boundedness of solutions to the vector equation (2.14) derived from a condition set for the scalar linear equation (5.2).

**Theorem 7**. Assume that the conditions of Theorem 1 and Lemma 7 are satisfied, and both $F_0$ and $\|\varphi\|$ are sufficiently small. Then,

$$|x(t,\varphi)| \le y(t,\phi) \le u(t,\phi), \ \forall t \ge t_0 \tag{5.8}$$

where $x(t,\varphi)$, $y(t,\phi)$, and $u(t,\phi)$ are the solutions to equations (2.14), (3.10), and (5.2), respectively, and $\phi(t) = |\varphi(t)|, \ \forall t \in [t_0 - \bar{h}, t_0]$.

**Proof.** In fact, applying Lemma 7 implies that inequality (5.1) is held under the stated conditions, which also justifies the application of Lemma 4 to equations (3.10) and (5.2). This, in turn, ensures that inequality (5.8) holds □

Thus, (5.7) and (5.8) give that:

$$|x(t,\varphi)| \le u_h(t, |\varphi(t)|) + F_0 u_{nh}(t) \tag{5.9}$$

This section introduces simplified linear auxiliary equations that offer more conservative, yet easier-to-apply, criteria for assessing the boundedness and stability of systems (2.14) and (2.15). However, a more accurate characterization of the dynamics of the original vector systems can be obtained through the direct simulations of their nonlinear scalar counterparts (3.10) and (3.12). These simulations serve to validate the methodology and assess its accuracy.

**6. Simulations**

Let us assume that (2.14) takes the following form,

$$\begin{aligned} &\dot{x} = (A_0(t) + A_1(t))x + \rho A_1(t) x(t-h) + f(t, x(t-h)) + F(t), \ \forall t \ge t_0 = 0 \\ &x(t) \in \mathbb{R}^2, x(t) = x_0, \ \forall t \in [-h, 0] \end{aligned} \tag{6.1}$$

where $f(t, x(t-h)) = b\left[0, \ x_2^3(t-h)\right]^T$, $F(t) = [0, F_2]^T$, $b, h, \rho \in \mathbb{R}_+$, $x_0 \in \mathbb{R}^2$ is a constant vector,

$$A_0(t) = \begin{pmatrix} \lambda(t) & 0 \\ 0 & \lambda(t) \end{pmatrix}, \ A_1(t) = \begin{pmatrix} 0 & 1 \\ -\omega(t) & -\alpha_1 \end{pmatrix}.$$

Next, we calculate the functions $p(t)$ and $c(t)$ for the subsystem $\dot{x} = A_0(t)x$ (see Section 3). Clearly, in this case, $w(t) = diag(\eta, \eta)$, where $\eta = \exp\left(\int_{t_0}^{t} \lambda(s)\, ds\right)$, which implies that $p(t) = \lambda(t)$ and $c(t) \equiv 1$. Thus, the scalar auxiliary equation (3.10) for (6.1) takes the following form:

$$\begin{aligned} &\dot{y} = (\lambda(t) + |A_1(t)|) y + \rho |A_1(t)| y(t-h) + b y^3(t-h) + |F(t)|, \ \forall t \ge 0, \ y(t) \in \mathbb{R}_{\ge 0} \\ &y(t) = |x_0|, \ \forall t \in [-h, 0] \end{aligned} \tag{6.2}$$

In turn, the autonomous scalar counterpart of (6.2) can be written as follows:

$$\begin{aligned} &\dot{\hat{y}} = (\hat{\lambda} + \hat{A}_1)\hat{y} + \rho \hat{A}_1 \hat{y}(t-h) + b\hat{y}^3(t-h) + F_0, \ \forall t \ge 0, \ \hat{y}(t) \in \mathbb{R}_{\ge 0} \\ &\hat{y}(t) = |x_0|, \ \forall t \in [-h, 0] \end{aligned} \tag{6.3}$$

where $\sup_{\forall t \ge t_0} \lambda(t) = \hat{\lambda} < 0, \ \sup_{\forall t \ge t_0} |A_1(t)| = \|A_1\| = \hat{A}_1$. Thus, due to Lemma 4, we infer that $y(t) \le \hat{y}(t), \ \forall t \ge 0$.

Recall that the stability of the linear equation $\dot{\hat{y}} = (\hat{\lambda} + \hat{A}_1)\hat{y} + \rho \hat{A}_1 \hat{y}(t-h)$ is well studied within the space of its parameters, and its asymptotic stability implies the asymptotic stability of the trivial solution of the homogeneous counterpart to equation (6.3), provided that $|b|$ is sufficiently small [30]. Moreover, the right sides of equations (6.2) and (6.3) depend only on the scalar function $|A_1(t)|$ and scalar parameter $\hat{A}_1$, respectively, rather than on all four entries of matrix $A_1(t)$. This significantly simplifies the sensitivity analysis of the original equation (6.1) with respect to the parameter variations.

To describe the results of simulations of equations (6.1)-(6.3), we set below that $\lambda = \lambda_0 + \lambda_+(t)$, where $\lambda_0 = -3$ and either a) $\lambda_+(t) = q\sin(dt),\ q = 0.1,\ d = 5$ or b) $\lambda_+(t) = q\exp(-dt),\ q = 1,\ d = 1$. Note that in the former case $\hat{\lambda} = -2.9$ whereas in the later one $\hat{\lambda} = -2$. Next, we also assume that $\omega(t) = \omega_0 + \omega_1(t)$, $\omega_1(t) = a_1 \sin(t) + a_2 \sin(\pi t)$, where $\omega_0 = 1,\ a_1 = a_2 = 0.1$. Additionally, we set in our simulations that $e(t) = [0, \sin(10t)]^T$; the values of other parameters are indicated in the captions of the figures. All simulations were caried out using the MATLAB solver DDE23.

Figure 1 contrasts the temporal progression of the solution norms for equations (6.1), (6.2), and (6.3), simulated with matched history functions and shown as solid, dashed, and dash-dotted lines, respectively. In all plots, the inequality $|x(t, x_0)| \le y(t, |x_0|) \le \hat{y}(t, |x_0|)$ holds over the simulated time intervals and across a wide range of system parameters, supporting our theoretical inferences. The solutions to the scalar nonautonomous equation (6.2) provide fairly accurate approximations of the norms of the corresponding solutions to (6.1). As expected, the simulations based on autonomous equation (6.3) yield cruder estimates. The accuracies of these estimates are more sensitive to increases in the delay parameter $h$, magnitude of the delay nonlinear term $b$, and amplitude of input $F_0$. Nevertheless, equation (6.2) continues to deliver adequate norm estimates for equation (6.1) across the tested parameter ranges.

<table>
<tr>
<td>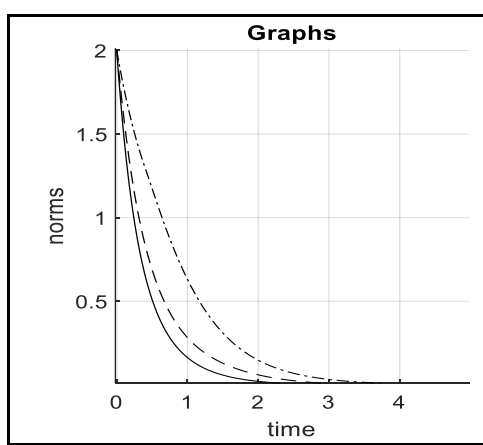
</td>
<td>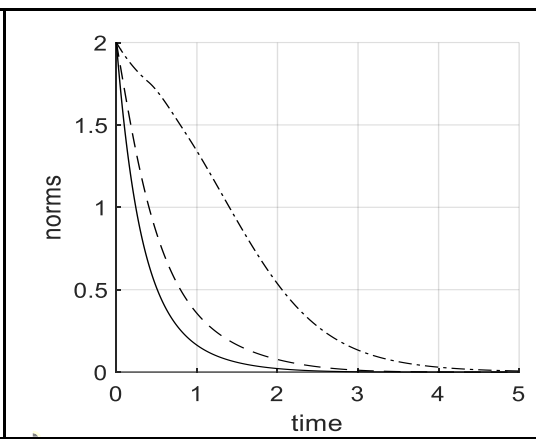
</td>
<td>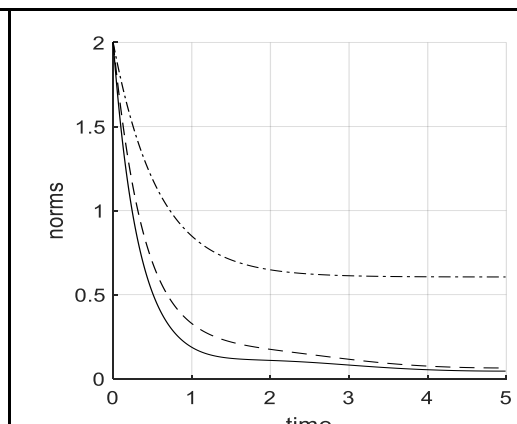
</td>
<td>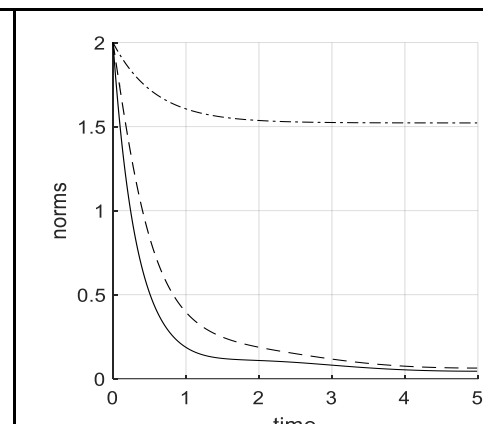
</td>
<td>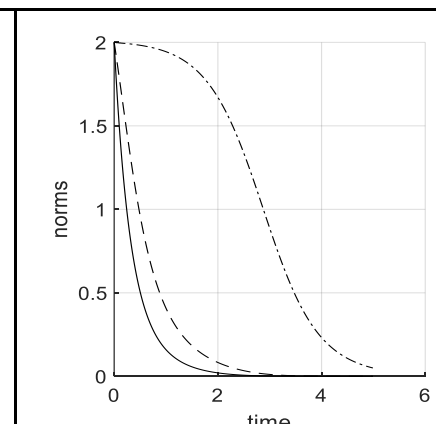
</td>
</tr>
<tr>
<td>Figure 1a. Case a. $h = 0.4, b = 0.1, F_0 = 0$</td>
<td>Figure 1b. Case a. $h = 0.4,\ b = 0.3, F_0 = 0$</td>
<td>Figure 1c. Case a. $h = 15, b = 0.1, F_0 = 0$</td>
<td>Figure 1d. Case a. $h = 15, b = 0.3, F_0 = 0$</td>
<td>Figure 1e. Case a. $h = 0.2, b = 0.4, F_0 = 0$</td>
</tr>
<tr>
<td>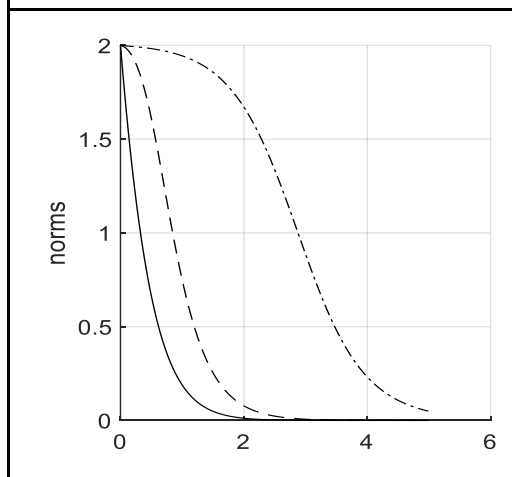
</td>
<td>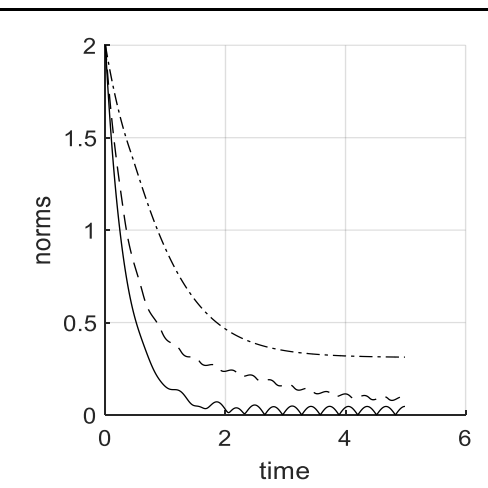
</td>
<td>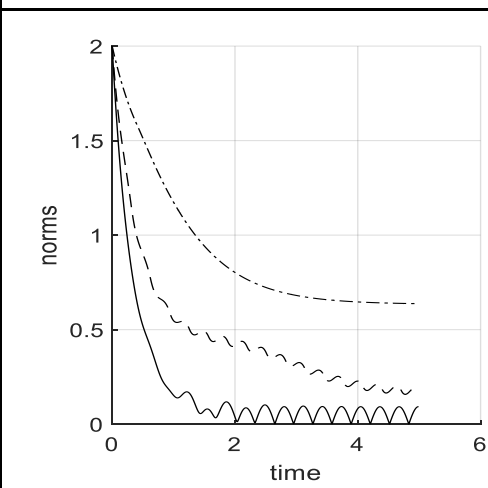
</td>
<td>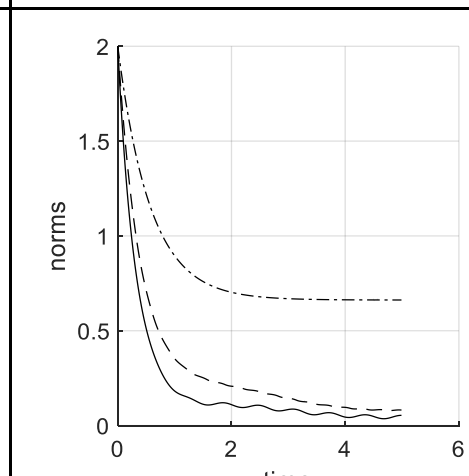
</td>
<td>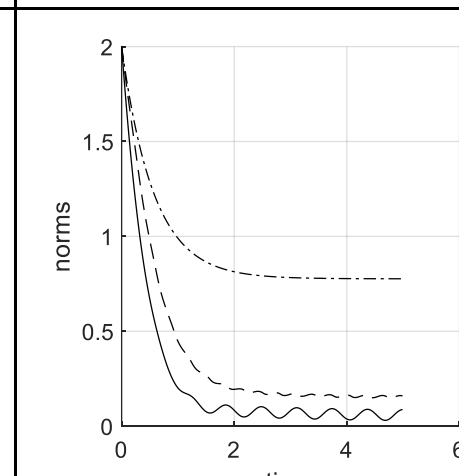
</td>
</tr>
<tr>
<td>Figure 1f. Case, b. $h = 0.2, b = 0.4, F_0 = 0$</td>
<td>Figure 1g. Case, a. $h = 0.2, b = 0.1, F_0 = 0.3$</td>
<td>Figure 1h. Case, a. $h = 0.2, b = 0.4, F_0 = 1$</td>
<td>Figure 1i. Case, a. $h = 0.2, b = 0.4, F_0 = 0$</td>
<td>Figure 1j. Case, b. $h = 15, b = 0.1, F_0 = 0.3$</td>
</tr>
</table>

**Figure 1**. *Comparison of the time evolution of the solution norms for vector equation (6.1) and the corresponding scalar equations (6.2) and (6.3)*

To simulate the boundaries of the trapping and stability regions, defined as sets of initial vectors that lead to bounded or stable solutions of equation (6.1) or its homogeneous counterpart, we represent the initial vector in polar coordinates as $x_{0,1} = r_0 \cos\varphi_0$, $x_{0,2} = r_0 \sin\varphi_0$. The angle coordinate was then discretized with a step size equal to $\pi / 100$. For each angle, the radial coordinate was adjusted incrementally to approximate the boundary point of the region. A binary search was used in the simulations, starting from a small radial value and increasing it until rapid growth in $|x(t, x_0)|$ was observed, indicating that the boundary had been crossed.

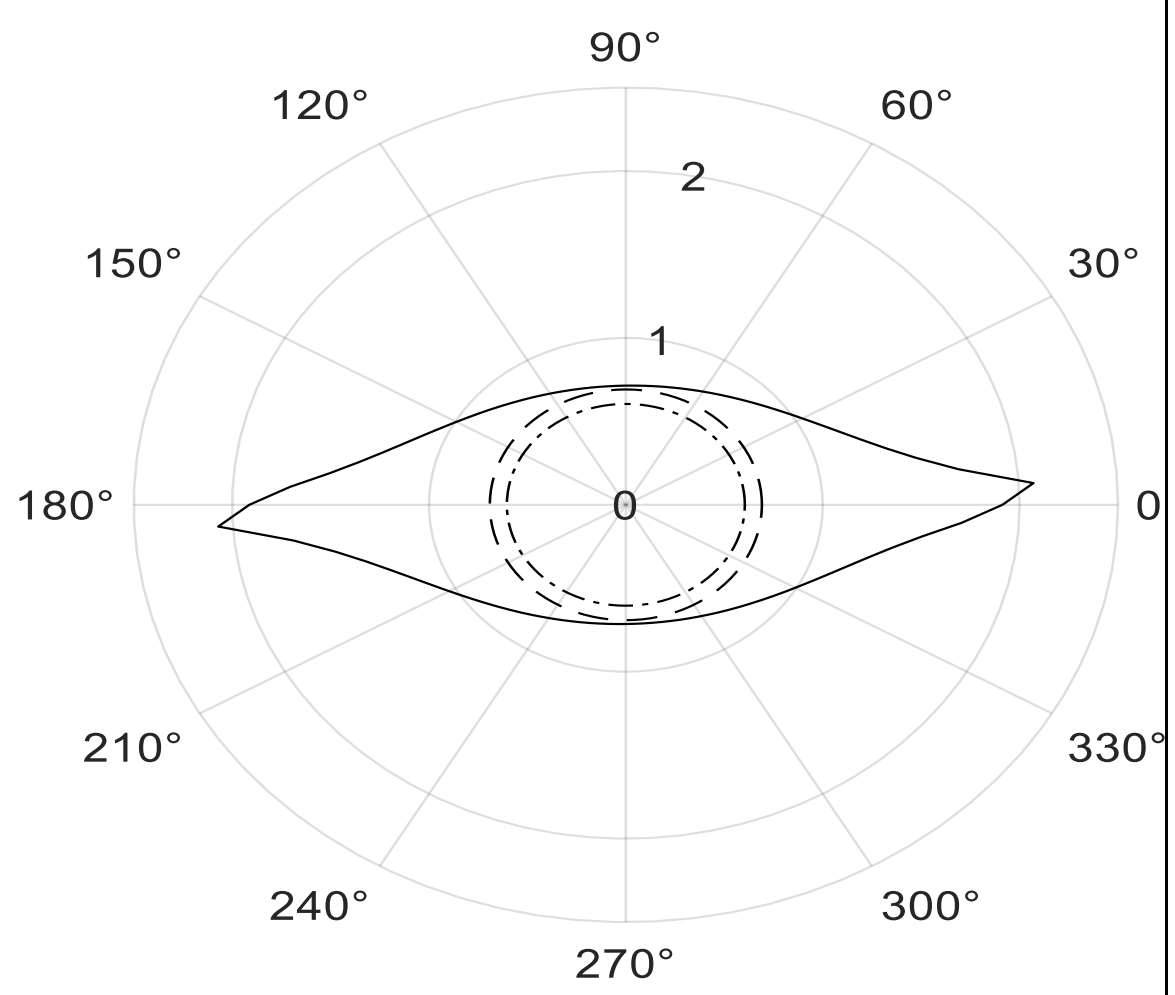

**Figure 2a**. Estimation of stability region. Case, a. $h = 0.4,\ b = 0.1,\ F_0 = 0$

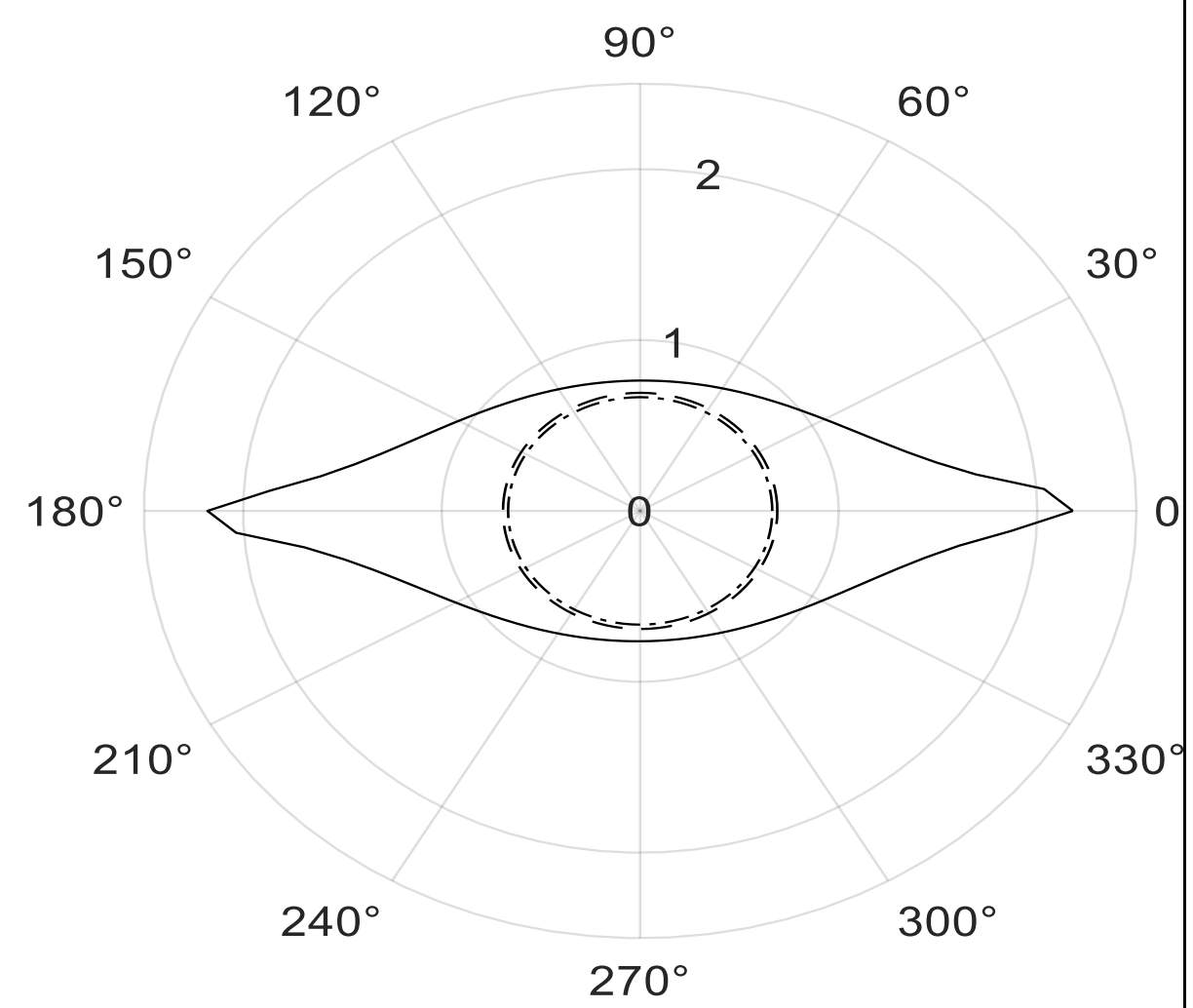

**Figure 2b**. Estimation of stability region. Case, a. $h = 15,\ b = 0.1,\ F_0 = 0$

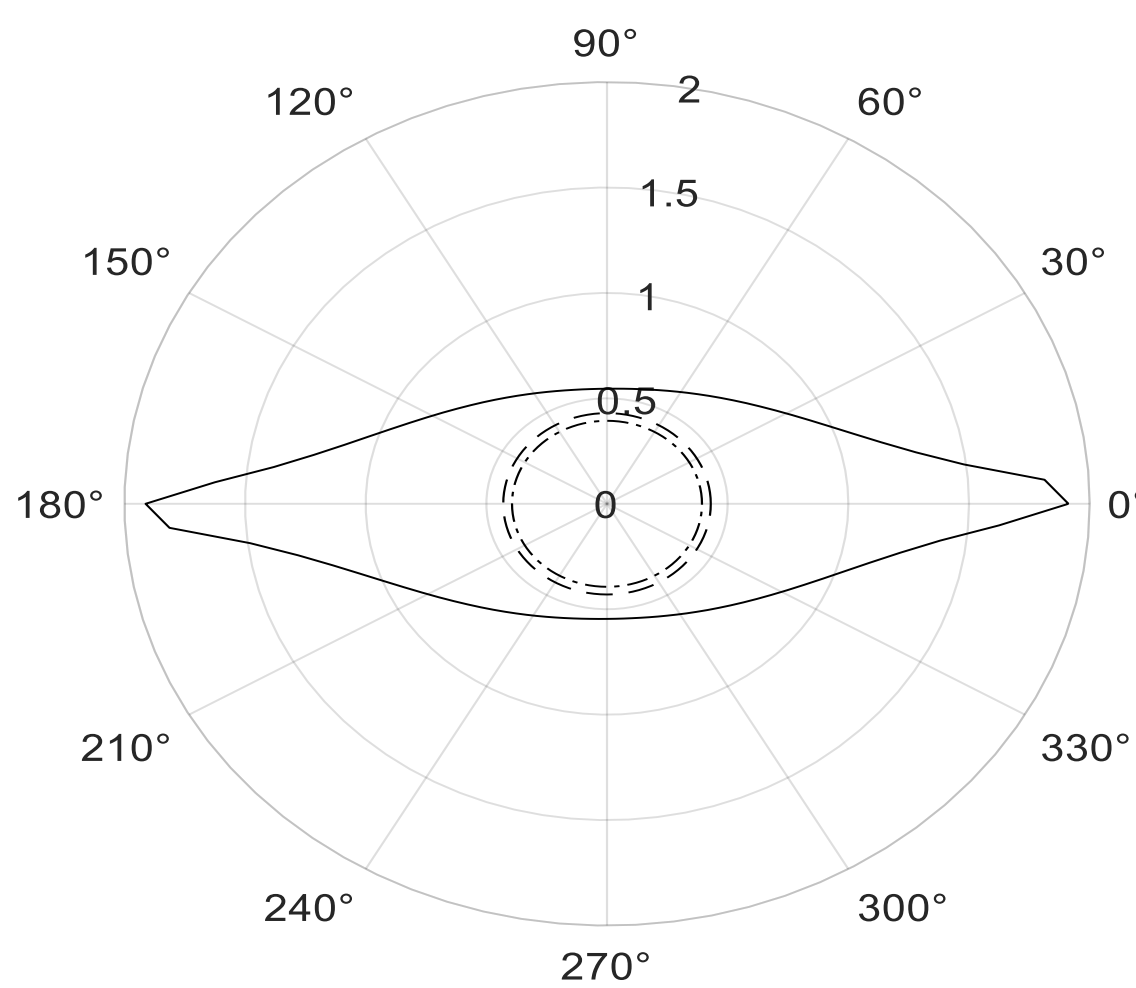

**Figure 2c**. Estimation of stability region. Case, b. $h = 15,\ b = 0.3,\ F_0 = 0$

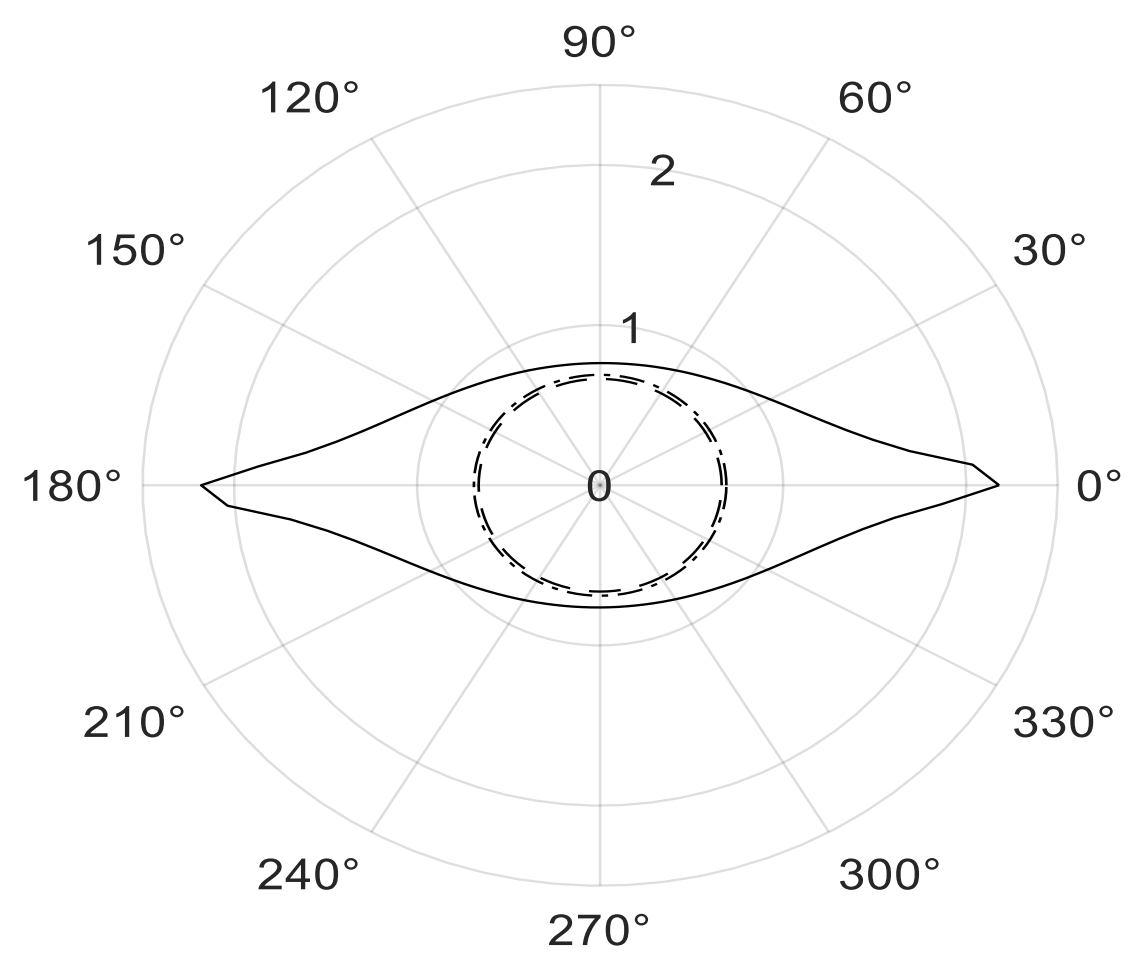

**Figure 2d**. Estimation of boundedness region. Case, a. $h = 15,\ b = 0.4,\ F_0 = 0.1$

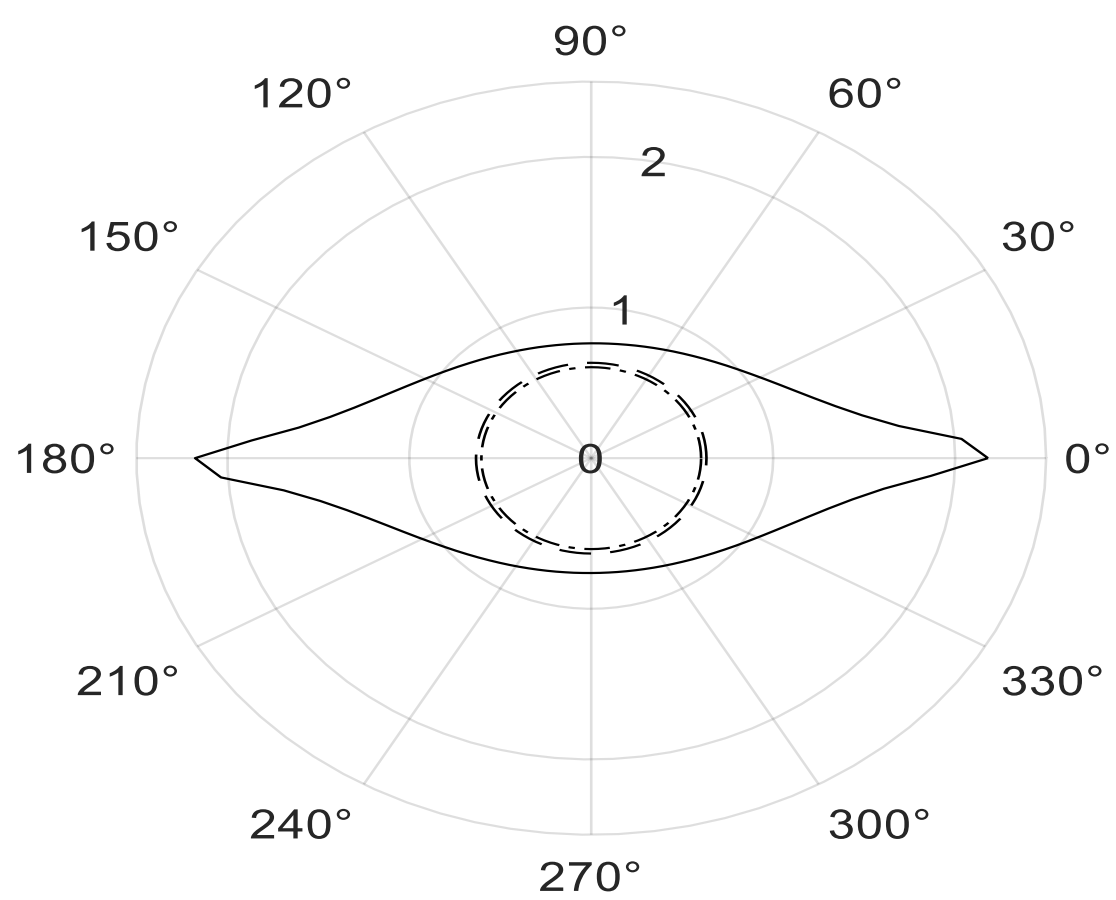

**Figure 2e**. Estimation of boundedness region. Case, a. $h=15,\ b=0.1,\ F_0=3$

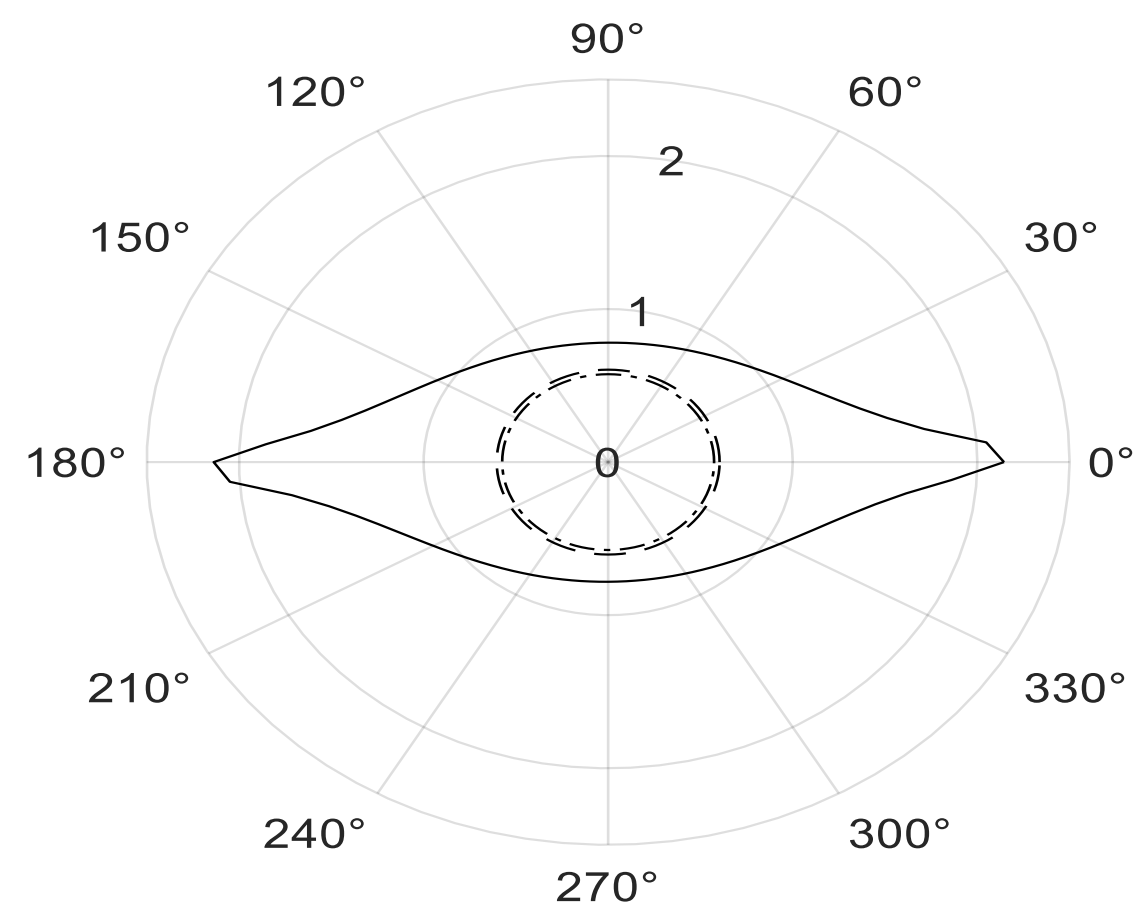

**Figure 2f**. Estimation of boundedness region. Case, b. $h=15,\ b=0.1,\ F_0=3$

**Figure 2**. *Estimated radii of disks embedded within the trapping* ( $F_0 \neq 0$ ) *or stability* ( $F_0 = 0$ ) *regions for equation (6.1) and its homogeneous counterpart.*

A similar approach was applied to scalar equations (6.2) and (6.3) to approximate the radii of the disks embedded within the trapping or stability regions of equation (6.1) and its homogeneous counterpart. The computation time required to estimate these radii is practically unaffected by the number of equations coupled in the original vector system. It is further reduced due to monotonic increase in both $y\left(t,|x_0|\right)$ and $\hat{y}\left(t,|x_0|\right)$, $\forall t \geq t_0$ with respect to $|x_0|$. In contrast, estimating the boundary of the trapping or stability regions for the vector equation (2.14) scales with $O\left(m^n\right)$, where $m$ is the number of discretization points per phase-space variable and $n$ is the number of such variables.

Figure 2 displays the boundaries of the trapping ( $F_0 \neq 0$ ) and stability ( $F_0 = 0$ ) regions obtained from the simulations of equations (6.1), (6.2), and (6.3), shown as solid, dashed, and dashed dotted lines, respectively, for the different parameter sets indicated in the figure captions. The curves are plotted in polar coordinates, with the natural logarithm of the radial coordinate used instead of the actual radius.

Thus, simulations of equations (6.2) and (6.3) enable efficient estimation of the radii of the disks embedded within the trapping and stability regions of equation (6.1) and its homogeneous counterpart. Across all plots in Figure 2, these estimates remained consistently close and aligned well with the smallest dimension of the actual trapping and stability regions. The estimated radii showed low sensitivity to variations in the delay parameter, input amplitude, and scale of the nonlinear term. The consistent arrangement of the curves across all plots in Figure 2 further supports our theoretical findings.

## 7. **Conclusion**

This study introduces a novel framework for analyzing a broad class of vector nonlinear systems with variable coefficients and multiple time-varying delays. The approach constructs scalar auxiliary equations whose solutions serve as upper bounds for the norms of the solutions to the original vector systems, provided that the history functions of the scalar and vector equations are appropriately aligned. This approach enables the assessment of boundedness and stability of complex vector delay systems by analyzing their scalar counterparts via either simplified analytical methods or efficient and rigorously justified simulations.

Using this framework, we derived new boundedness and stability criteria, estimated the radii of balls containing history functions that lead to bounded or stable behavior, and simulated the upper bounds on the norms of solutions to the original systems.

In addition, using a Lipschitz-like continuity condition for the nonlinear terms, we derived linear counterparts for the scalar auxiliary equations. We show that the solutions of these linear scalar delay equations serve as upper bounds for their nonlinear counterparts, further simplifying the analysis and yielding additional boundedness and stability criteria.

An important advantage of the proposed method is its natural aggregation of parameters, history functions, and uncertainty models, which simplifies the sensitivity and robustness analyses.

Our results are validated through representative simulations, showing that the solutions to the scalar auxiliary equations fairly approximate the norms of the solutions to the original vector systems initiated from the central regions of the boundedness or stability domains. We also demonstrated that our method estimates the radii of the boundedness and stability balls with a relatively small error for a wide range of parameters of the test system.

However, this approach has some limitations. First, deriving the main scalar equation (3.10) involves numerically computing the fundamental solution matrix of the linear system (2.16), which may become computationally demanding for high-dimensional systems. Second, the repeated application of norm inequalities to nonlinear terms can introduce conservatism, particularly when these terms are relatively large. Lastly, the scalar function $c(t)$ may exhibit long-term growth, further contributing to conservative estimates. These issues will be addressed in our future studies under additional assumptions that preserve the practical applicability of the methodology.

Future work will also focus on extending the applicability of the methodology by developing efficient recursive approximations for the boundaries of trapping and stability regions as well as bilateral bounds on the norms of solutions to vector nonlinear and nonautonomous delay systems.

**Appendix**. Let us demonstrate how to derive inequality (3.3) using a simple example that can be naturally extended to more complex cases. Assume that $x=[x_1\, x_2]^T$ and a vector function $f$ is defined as follows $f=\left[a_1(t)x_1^3x_2^2(t-h_1) \quad a_2(t)x_2^3(t-h_2)\right]^T$ . Then $|f|_2 \le |f|_1 \le |a_1||x_1^3||x_2^2(t-h_1)| + |a_2||x_2^3(t-h_2)|$ $\le |a_1||x(t)|^3|x(t-h_1)|^2+|a_2||x(t-h_2)|^3$, where we use that $|x_i^n| \le |x|^n$, $|x_i^n(t-h)| \le |x(t-h)|^n$, $i=1,2,\ n\in\mathbb{N}$ . Clearly, this inference can be extended to the power series and some rational functions under the obtained conditions.

**Acknowledgement**. The code for simulations discussed in Section 6 was developed by Steve Koblik.

Data sharing is not applicable to this article, as no datasets were generated or analyzed during the current study.

The author declares that he has no conflict of interest.